\documentclass[a4paper,10pt,twoside,bibnote]{cpc-hepnp}
\usepackage{CJK,upgreek,fancyhdr}
\usepackage{multicol}
\usepackage{graphicx}
\usepackage{booktabs}
\usepackage{slashed}
\usepackage{amssymb,bm,mathrsfs,bbm,amscd}
\usepackage[tbtags]{amsmath}
\usepackage{lastpage}
\usepackage{multirow}
\usepackage{textcomp}
\usepackage{epstopdf}
\usepackage{float}
\usepackage{footnote}

\let\jnfont=\rm
\def\NPB#1,{{\jnfont Nucl.\ Phys.\ B, }{\bf #1}:}
\def\PLB#1,{{\jnfont Phys.\ Lett.\ B, }{\bf #1}:}
\def\EPJC#1,{{\jnfont Eur.\ Phys.\ Jour.\ C, }{\bf #1}:}
\def\PRD#1,{{\jnfont Phys.\ Rev.\ D, }{\bf #1}:}
\def\PRL#1,{{\jnfont Phys.\ Rev.\ Lett.,\ }{\bf #1}:}
\def\MPLA#1,{{\jnfont Mod.\ Phys.\ Lett.\ A, }{\bf #1}:}
\def\JPG#1,{{\jnfont J.\ Phys.\ G, }{\bf #1}:}
\def\CTP#1,{{\jnfont Commun.\ Theor.\ Phys.,\ }{\bf #1}:}
\def\ZPC#1,{{\jnfont Z.\ Phys.,\ C }{\bf #1}:}
\def\JHEP#1,{{\jnfont JHEP, \ }{\bf #1}:}
\def\lsim{\raise0.3ex\hbox{$<$\kern-0.75em\raise-1.1ex\hbox{$\sim$}}}
\def\gsim{\raise0.3ex\hbox{$>$\kern-0.75em\raise-1.1ex\hbox{$\sim$}}}

\newcommand{\SLASH}[2]{\makebox[#2ex][l]{$#1$}/}
\newcommand{\pslash}{\SLASH{p}{.2}}
\newcommand{\GeV}{~{\rm GeV}}
\newcommand{\TeV}{~\rm TeV}
\newcommand{\fbm}{{~\rm fb}^{-1}}

\begin{document}

\setlength{\abovecaptionskip}{4pt plus1pt minus1pt}
\setlength{\belowcaptionskip}{4pt plus1pt minus1pt}
\setlength{\abovedisplayskip}{6pt plus1pt minus1pt}
\setlength{\belowdisplayskip}{6pt plus1pt minus1pt}
\addtolength{\thinmuskip}{-1mu}
\addtolength{\medmuskip}{-2mu}
\addtolength{\thickmuskip}{-2mu}
\setlength{\belowrulesep}{0pt}
\setlength{\aboverulesep}{0pt}
\setlength{\arraycolsep}{2pt}

\fancyhead[c]{\small Chinese Physics C~Vol. 44, No. 06 (2020) 061001 ~~~~~~~~~~~~~~~~~~~~~~WHU-HEP-PH-TEV005~~~~~~~~~~~~~~~~~~~arXiv:1911.08319}
\fancyfoot[C]{\small 061001-\thepage}
\footnotetext[0]{Published online 27 April 2020}

\title{The semi-constrained NMSSM in light of muon g-2, LHC, and\\ dark matter constraints\thanks{Supported by
National Natural Science Foundation of China (NNSFC) (11605123,
11675147, 11547103, 11547310),  the Innovation Talent project of
Henan Province (15HASTIT017), and the Young Core Instructor
Foundation of Henan Education Department. J. Z. also thanks the
support of the China Scholarship Council (CSC) (201706275160)
while at the University of Chicago
as a visiting scholar, and the U.S. National
Science Foundation (NSF) (PHY-0855561) while at Michigan State University from 2014-2015. }}

\title{The Light Higgsino-dominated NLSPs in the Semi-constrained NMSSM\thanks{supported by the National Natural Science Foundation of China (NNSFC) under Grant No. 11605123.}}

\author{
      Kun Wang$^{1;1)}$\email{wk2016@whu.edu.cn}
\quad Jingya Zhu$^{1;2)}$\email{zhujy@whu.edu.cn}
}
\maketitle

\address{
$^1$ Center for Theoretical Physics, School of Physics and Technology, Wuhan University, Wuhan 430072, China \\
}

\begin{abstract}
In the semi-constrained NMSSM (scNMSSM, or NMSSM with non-universal Higgs mass) under current constraints, we consider a scenario where $h_2$ is the SM-like Higgs, $\tilde{\chi}^0_1$ is singlino-dominated LSP, $\tilde{\chi}^{\pm}_1$ and $\tilde{\chi}^0_{2,3}$ are mass-degenerated, light and higgsino-dominated NLSPs (next-to-lightest supersymmetric particles).
We investigate the constraints to these NLSPs from searching for SUSY particles at the LHC Run-I and Run-II,
discuss the possibility of discovering these NLSPs in the future,
and come to the following conclusions:
(i) With all data of Run I and up to $36\fbm$ data of Run II at the LHC, the search results by ATLAS and CMS can still not exclude the higgsino-dominated NLSPs of $100\sim200\GeV$.
(ii) When the mass difference with $\tilde{\chi}^0_{1}$ is smaller than $m_{h_2}$, $\tilde{\chi}^0_{2}$ and $\tilde{\chi}^0_{3}$ have opposite preference on decaying to $Z/Z^*$ or $h_1$.
(iii) When the mass difference between NLSP and LSP is larger than $m_Z$, most of the samples can be checked at $5\sigma$ level with future $300\fbm$ data at the LHC.
While with $3000\fbm$ data at the High Luminosity LHC (HL-LHC), nearly all of the samples can be checked at $5\sigma$ level even if the mass difference is insufficient.
(iv) The $a_1$ funnel and the $h_2/Z$ funnel mechanisms for the singlino-dominated LSP annihilating can not be distinguished by searching for NLSPs.
\end{abstract}


\begin{pacs}
11.30 Pb     \qquad     {\bf DOI:} 10.1088/1674-1137/44/06/061001
\end{pacs}

\footnotetext[0]{\hspace*{-3mm}\raisebox{0.3ex}{$\scriptstyle\copyright$}2020
Chinese Physical Society and the Institute of High Energy Physics
of the Chinese Academy of Sciences and the Institute
of Modern Physics of the Chinese Academy of Sciences and IOP Publishing Ltd}%

\vspace{0.7mm}
\begin{multicols}{2}

\section{Introduction}
\label{sec:introduction}

As an internal symmetry between fermions and bosons, Supersymmetry (SUSY) is an attractive idea.
In the framework of SUSY, the strong, weak and hypercharge gauge couplings ($g_3,~ g_2,~ g_1$) can be unified at the GUT scale ($\sim 10^{16} \GeV$), and the large hierarchy problem between the electroweak scale and the Planck scale can be solved.
Besides, with the R-parity conserved, the lightest SUSY particle (LSP) is stable and can be a good candidate for weakly-interaction-massive-particle (WIMP) dark matter (DM).

SUSY at TeV scale is motivated by the possible cancellation of quadratic divergences of the Higgs boson mass.
And the simplest implementation of SUSY is the Minimal Supersymmetric extension to the Standard Model (MSSM).
Since the soft SUSY breaking parameters is totally free in the MSSM, a dynamic way to get these parameters is more favored.
In the minimal supergravity (mSUGRA), the K\"{a}ler potential is employed to yield the minimal kinetic energy terms for the MSSM fields,
where all the trilinear couplings, gaugino, and scalar mass parameters unify respectively at the GUT scale.
The fully constrained MSSM (CMSSM) is the MSSM with the boundary conditions the same as the mSUGRA.
But to get a 125 GeV SM-like Higgs, the MSSM needs very large one-loop radiative corrections to Higgs mass, which makes the MSSM not natural.
And there is a so-called $\mu$-problem \cite{Kim:1983dt} in the MSSM, where the superpotential contains a term $\mu \hat{H}_u \hat{H}_d$, and $\mu$ is the only dimensionful parameter and has to be chosen artificially.

The Next-to Minimal Supersymmetric Standard Model (NMSSM) can solve the $\mu$-problem by introducing a complex singlet superfield $\hat{S}$, which can generate an effective $\mu$-term dynamically.
And it can easily predict an SM-like 125 GeV Higgs, under all the constraints and with low fine-tuning \cite{125-susy}.
The fully constrained NMSSM (cNMSSM) contains none or only one more parameter than the CMSSM/mSUGRA, thus both of them are in tension with current experimental constraints including 125 GeV Higgs mass, high mass bound of gluino, muon g-2, and dark matter \cite{125-cNMSSM,cNMSSM-pre125, cmssm-125, cmssm-zy, cmssm-killing, gambit:CMSSM}.
So, we consider the semi-constrained NMSSM (scNMSSM), which relaxes the unification of scalar masses by decoupling the squared-masses of the Higgs bosons and the squarks/sleptons, which is also called NMSSM with non-universal Higgs mass  (NUHM) \cite{scNMSSM, Wang-scNMSSM, scnmssm scan}.
In the scNMSSM, the bino and wino are heavy because the high mass bound of gluino and the unification of gaugino masses at GUT scale, thus the light neutralinos and charginos can only be singlino-dominated or higgsino-dominated$^{1)}$.
\footnotetext{1) The behaviors of the cNMSSM under recent constraints is similar as the CMSSM, where the additional parameter $\lambda$ is very small, and higgsino mass parameter $\mu_{\rm eff}$ is calculated to be very large, thus the higgsino-dominated and singlino-dominated neutralinos are very heavy \cite{125-cNMSSM,cNMSSM-pre125}.}

In recent years, the ATLAS and CMS collaborations have carried out many searches for SUSY particles, which pushed the gluino and squarks masses bounds in simple models up to several hundred GeV and even TeV scale.
While it is still possible for the electroweakino sector to be very light.
The electroweakino sector of NMSSM was studied in \cite{singlino-dominated LSP in NMSSM, Pyarelal:2019zth, LHC search nmssm, nmssm signal 3l, nmssm signal multi-lepton, nmssm signal lowmet, singlino-LSP},
among which different search channels were provided, such as multi-leptons \cite{nmssm signal 3l, nmssm signal multi-lepton}, and jets with missing transverse momentum ($\pslash_T$) \cite{nmssm signal lowmet}.
These motivated us to check the current status of higgsino, in special SUSY models such as the scNMSSM, under direct-search constraints and their possibility of discovery by detailed simulation.

In this work, we discuss the light higgsino-dominated NLSPs (next-to-lightest supersymmetric particles) in the scNMSSM.
We use the scenario of singlino-dominated $\tilde{\chi}^0_1$ and SM-like $h_2$ in the scan result in our former work on scNMSSM \cite{Wang-scNMSSM}, where we considered the constraints including theoretical constraints of vacuum stability and Landau pole, experimental constraints of Higgs data, muon g-2, B physics, dark matter relic density and direct searches, etc.
Thus in this scenario the $\tilde{\chi}^{\pm}_1$ and $\tilde{\chi}^0_{2,3}$ are higgsino-dominated, light and mass-degenerated NLSPs.
We first investigate the constraints to these NLSPs, including searching for SUSY particle at the LHC Run-I and Run-II.
We use Monte Carlo to do the detailed simulations to impose these constraints.
Then we discuss the possibility of discovering the higgsino-dominated NLSPs in the future at the High Luminosity LHC (HL-LHC).

This paper is organized as follows.
First, in Section 2, we briefly introduce the model of NMSSM and scNMSSM, especially the Higgs and electroweakino sector.
Later in Section 3, we discuss the constraints to the light higgsino-dominated NLSPs, and the possibility of discovering them at the HL-LHC.
Finally, we draw our conclusions in Section 4.

\section{Introduction to the NMSSM and scNMSSM}
\label{sec:nmssm}

The superpotential of NMSSM with $\mathbb{Z}_3$ symmetry :
\begin{equation}
\label{eq:w_nmssm}
W_{\rm NMSSM} = W_{\rm MSSM}|_{\mu=0} + \lambda \hat{S} \hat{H}_u \cdot \hat{H}_d + \frac{\kappa}{3} \hat{S} ^3  \,,
\end{equation}
where the superfields $\hat{H}_u$ and $\hat{H}_d$ are two complex doublet superfields,
the superfield $\hat{S}$ is the singlet superfield,
the coupling constants $\lambda$ and $\kappa$ are dimensionless,
and the $W_{\rm MSSM}|_{\mu=0}$ is actually the Yukawa couplings of the $\hat{H}_u$ and $\hat{H}_d$ to the quark and lepton superfields.
When electroweak symmetry breaking, the scalar component of superfields $\hat{H}_u$ , $\hat{H}_d$ and $\hat{S}$ get their vacuum expectation values (VEVs) $v_u$, $v_d$ and $v_s$ respectively. The relations between the VEVs are
\begin{eqnarray}
\tan\beta = v_u / v_d, ~~
v= \sqrt{v_u^2+v_d^2}=174 \GeV, ~~
\mu_{\rm eff} = \lambda v_s,
\end{eqnarray}
where the $\mu_{\rm eff}$ is the mass scale of higgsino, like in the MSSM.
In the following, for the sake of convenience, we refer to $\mu_{\rm eff}$ as $\mu$.


The soft SUSY breaking terms in the NMSSM is only different from the MSSM in several terms:
\begin{eqnarray}
\label{eq:soft}
-\mathcal{L}_{\rm NMSSM}^{\rm soft} &=& -\mathcal{L}_{\rm MSSM}^{\rm soft}|_{\mu=0} + {m}_S^2 |S|^2
+ \lambda A_\lambda S H_u \cdot H_d \nonumber\\
&&+ \frac{1}{3} \kappa A_\kappa S^3 + h.c. \,,
\end{eqnarray}
where the $S$, $H_u$ and $H_d$ is the scalar component of the superfields,
the ${m}_S^2$ is the soft SUSY breaking mass for singlet field $S$,
and the trilinear coupling constants $A_\lambda$ and $A_\kappa$ have mass dimension.

In the semi-constrained NMSSM (scNMSSM), the Higgs sector are considered non-universal, that is, the Higgs soft mass $m^2_{H_u}$,$m^2_{H_d}$ and $m^2_{S}$ are allowed to be different from $M^2_0+\mu^2$, and the trilinear couplings $A_\lambda$, $A_\kappa$ can be different from $A_0$.
Hence, in the scNMSSM, the complete parameter sector is usually chosen as:
\begin{equation}
\label{eq:parameter}
\lambda,\,\, \kappa,\,\, \tan\beta,\,\, \mu,\,\, A_\lambda,\,\, A_\kappa,\,\, A_0,\,\, M_{1/2}, \,\,M_0  \, .
\end{equation}

\subsection{The Higgs sector of NMSSM and scNMSSM}

When the electroweak symmetry breaking, the scalar component of superfields $\hat{H}_u$ , $\hat{H}_d$ and $\hat{S}$ can be written as
\begin{equation}
\label{eq:higgsfields}
H_u=\left(
\begin{array}{c}
H^+_u \\
v_u+\frac{ H^R_u + i H^I_u}{ \sqrt{2} }
\end{array} \right), ~
H_d=\left(
\begin{array}{c}
v_d+\frac{H^R_d + i H^I_d }{ \sqrt{2} }\\
H^-_d
\end{array} \right), ~
S = v_s + \frac{S^R+i S^I}{ \sqrt{2} } ,
\end{equation}
where $H^R_u$, $H^R_d$, and $S^R$ are CP-even component fields, $H^I_u$, $H^I_u$,
and $S^I$ are the CP-odd component fields,
and the $H^+_u$ and $H^-_d$ are charged component fields

In the basis $(H^R_d, H^R_u, S^R)$, the CP-even scalar mass matrix is \cite{Ellwanger nmssm}
\begin{equation}
\label{eq:l-higgs-s}
\mathcal{L} \ni \frac{1}{2}  \left( H^R_d, H^R_u, S^R \right) M_{S}^2
\left( \begin{array}{c}
H^R_d  \\
H^R_u \\
S^R
\end{array} \right)
\end{equation}
with
\begin{eqnarray}
\label{higgs-s-mass}
M_{S}^2= \left( \begin{array}{ccc}
M_A^2 s^2_\beta + M_Z^2 c^2_\beta            & (2 \lambda v^2 -M_A^2- M_Z^2)s_\beta c_\beta        & C c_{\beta}+ C' s_\beta   \\
(2 \lambda v^2 -M_A^2- M_Z^2)s_\beta c_\beta   & M_A^2 c^2_\beta + M_Z^2 s^2_\beta          &  C s_\beta + C' c_{\beta}  \\
C c_{\beta}+ C' s_\beta & C s_\beta + C' c_{\beta}  & M_{S,S^R S^R}^2
\end{array} \right)
\end{eqnarray}
where
\begin{eqnarray}
&& M_A^2 =   \frac{2\mu (A_\lambda + \kappa v_s)}{\sin 2\beta}, \\
&& C=2 \lambda^2 v v_s, ~~~~ C'=\lambda v (A_\lambda -2\kappa v_s), \\
&& \label{mass-sr} M_{S,S^R S^R}^2 =\lambda A_\lambda \frac{v_u v_d}{v_s} +\kappa v_s (A_\kappa + 4\kappa v_s),
\end{eqnarray}
and  $s_{\beta}=\sin\beta, c_{\beta}=\cos\beta$.
Actually, there is a more common basis $(H_1, H_2, S^R)$, where
\begin{eqnarray}
    H_1=H_u^R c_\beta - H_d^R s_\beta ,~~~~
    H_2=H_u^R s_\beta + H_d^R c_\beta
\end{eqnarray}
and the $H_2$ is the SM Higgs field.
In the basis $(H_1, H_2, S^R)$, the scalar mass matrix is different from Eq.(\ref{higgs-s-mass}).
But, since the rotation of the basis do not touch the third component $S^R$, the $M_{S, S^R S^R}^2$ will keep the same as in Eq.(\ref{mass-sr}).
The Higgs boson mass matrix $M_{S'}^2$ in basis  $(H_1, H_2, S^R)$ is given by \cite{NMSSM-Hmass}
\begin{align}
M_{S',H_1 H_1}^2 &=  M_A^2 + \left(m_Z^2- \lambda^2 v^2 \right)\sin^2 2 \beta \,, \\
M_{S',H_1 H_2}^2 &=  - \frac{1}{2}\left(m_Z^2- \lambda^2 v^2\right) \sin 4 \beta \,, \\
M_{S',H_1 S^R}^2 &=  -\left( \frac{M_A^2}{2 \mu /\sin 2 \beta } + \kappa v_s \right) \lambda v \cos 2 \beta \,,  \\
M_{S',H_2 H_2}^2 &=  m_Z^2 \cos^2 2 \beta + \lambda^2 v^2 \sin^2 2 \beta \,,  \\
M_{S',H_2 S^R}^2 &=  2 \lambda \mu v \left[1-\left( \frac{M_A}{2\mu / \sin 2 \beta} \right)^2 - \frac{\kappa}{2 \lambda } \sin 2 \beta  \right] \,,  \\
\label{msp} M_{S',S^R S^R}^2 &= \frac{1}{4} \lambda^2 v^2 \left(\frac{M_A}{\mu / \sin 2 \beta }\right)^2 + \kappa v_s A_\kappa +4(\kappa v_s)^2- \frac{1}{2} \lambda \kappa v^2 \sin 2 \beta \,.
\end{align}
And comparing Eq.(\ref{msp}) with Eq.(\ref{mass-sr}), it's not hard to get $M_{S',S^R S^R}^2=M_{S,S^R S^R}^2$.

To get the physical CP-odd scalar Higgs bosons, one can rotate the Higgs fields,
\begin{equation}
    A=H_u^I c_\beta  +H_d^I s_\beta ~.
\end{equation}
Then the Goldstone mode can be dropped off, and the CP-odd scalar mass matrix in the basis $(A, S^I)$ become \cite{Ellwanger nmssm}
\begin{equation}
\label{eq:l-higgs-p}
\mathcal{L} \ni \frac{1}{2}  \left( A, S^I \right) M_{P}^2
\left( \begin{array}{c}
A  \\
S^I
\end{array} \right)
\end{equation}
with
\begin{eqnarray}
\label{higgs-p-mass}
M_{P}^2= \left( \begin{array}{ccc}
M_A^2   & \lambda v (A_\lambda-2\kappa v_s)          \\
\lambda v (A_\lambda-2\kappa v_s)    & M_{P,S^I S^I}^2
\end{array} \right)
\end{eqnarray}
where
\begin{equation}
\label{mass-si}
    M_{P,S^I S^I}^2 = \lambda (A_\lambda+4\kappa v_s)  \frac{v_u v_d}{v_s}  - 3 \kappa v_s A_\kappa ~.
\end{equation}

The mass eigenstates of the CP-even Higgs $h_i$ ($i=1,2,3$) and the CP-odd Higgs $A_i$($i=1,2$) can be obtained by
\begin{equation}
\left( \begin{array}{c}
h_1  \\
h_2  \\
h_3
\end{array} \right)
= S_{ij}
\left( \begin{array}{c}
H_1  \\
H_2  \\
S^R
\end{array} \right) \,,
\qquad \quad
\left( \begin{array}{c}
a_1  \\
a_2
\end{array} \right)
= P_{ij}
\left( \begin{array}{c}
A  \\
S^I
\end{array} \right) \,,
\end{equation}
where the matrix $S_{ij}$ can diagonalize the mass matrix $M_{S'}^2$, and the matrix $P_{ij}$ can diagonalize the mass matrix $M_{P}^2$.

\subsection{The electroweakino sector of NMSSM and scNMSSM}

In the NMSSM, there are five neutralinos $\tilde{\chi}^0_{i}$ ($i=1,2,3,4,5$), which are the mixture of $\tilde{B}$ (bino), $\tilde{W}^{3}$ (wino), $\tilde{H}_{d}^0$, $\tilde{H}_{u}^0$ (higgsinos) and $\tilde{S}$ (singlino).
In the gauge-eigenstate basis
$\psi^0 = (\tilde{B}, \tilde{W}^{3}, \tilde{H}_{d}^0, \tilde{H}_{u}^0, \tilde{S})$,
the neutralino mass matrix takes the form \cite{Ellwanger nmssm}
\begin{eqnarray}
\label{Neu-mass}
M_{\tilde{\chi}^{0}}= \left( \begin{array}{ccccc}
    M_{1}        & 0                 & -c_{\beta} s_W m_Z  &  s_{\beta} s_W m_Z  & 0 \\
    0            & M_{2}             &  c_{\beta} c_w m_z  & -s_{\beta} c_W m_Z  & 0 \\
-c_{\beta} s_W m_Z &  c_{\beta} c_w m_z  & 0               & -\mu             & -\lambda v_{d} \\
 s_{\beta} s_W m_Z & -s_{\beta} c_W m_Z  & -\mu            & 0                & -\lambda v_{u} \\
    0            & 0                 & -\lambda v_{d}      & -\lambda v_{u}   & 2\kappa v_s \\
\end{array} \right)
\end{eqnarray}
where $s_{\beta}=\sin\beta, c_{\beta}=\cos\beta, s_W=\sin\theta_W, c_W=\cos\theta_W$. To get the mass eigenstates, one can diagonalize the neutralino mass matrix $M_{\tilde{\chi}^{0}}$
\begin{equation}
N^* M_{\tilde{\chi}^{0}} N^{-1} = M_{\tilde{\chi}^{0}}^{D}={\rm Diag}( m_{\tilde{\chi}^0_{\rm 1}},m_{\tilde{\chi}^0_{\rm 2}},m_{\tilde{\chi}^0_{\rm 3}},m_{\tilde{\chi}^0_{\rm 4}},m_{\tilde{\chi}^0_{\rm 5}})
\end{equation}
where $M_{\tilde{\chi}^{0}}^{D}$ means the diagonal mass matrix, and the order of eigenvalues is $m_{\tilde{\chi}^0_{\rm 1}}<m_{\tilde{\chi}^0_{\rm 2}}<m_{\tilde{\chi}^0_{\rm 3}}<m_{\tilde{\chi}^0_{\rm 4}}<m_{\tilde{\chi}^0_{\rm 5}}$. Meanwhile, one can get the mass eigenstates
\begin{eqnarray}
\left(  \begin{array}{c}
    \tilde{\chi}^0_{\rm 1} \\ \tilde{\chi}^0_{\rm 2} \\
    \tilde{\chi}^0_{\rm 3} \\ \tilde{\chi}^0_{\rm 4} \\
    \tilde{\chi}^0_{\rm 5} \\
  \end{array} \right)
=N_{ij} \left(  \begin{array}{c}
    \tilde{B} \\     \tilde{W^{0}} \\
    \tilde{H_{d}} \\    \tilde{H_{u}} \\
    \tilde{S} \\ \end{array} \right)
\end{eqnarray}

In the scNMSSM, bino and wino were constrained to be very heavy, because of the high mass bounds of gluino and the universal gaugino mass at the GUT scale, thus they can be decoupled from the light sector.
Then the following relations for the $N_{ij}$ can be found \cite{NIJ}:
\begin{equation}
    N_{i3}:N_{i4}:N_{i5}=
    \left[ \frac{m_{\tilde{\chi}^0_{\rm i}}}{\mu} s_\beta - c_\beta\right]
    : \left[ \frac{m_{\tilde{\chi}^0_{\rm i}}}{\mu} c_\beta - s_\beta\right]
    : \frac{\mu - m_{\tilde{\chi}^0_{\rm i}}}{\lambda v}
\end{equation}

We assume the lightest neutralino $ \tilde{\chi}^0_{\rm 1}$ is the lightest supersymmetric particle (LSP) and makes up of the cosmic dark matter.
If the LSP $\tilde{\chi}^0_{\rm 1}$ satisfies $N_{15}^2 > 0.5$, we call it singlino-dominated.
And the coupling of such an LSP with the CP-even Higgs bosons is given by \cite{NIJ}
\begin{eqnarray}
    C_{h_i \tilde{\chi}^0_{\rm 1} \tilde{\chi}^0_{\rm 1}} = \sqrt{2} \lambda \big[
    S_{i1}N_{15}(N_{13} c_\beta - N_{14} s_\beta ) +
       S_{i2}N_{15} ~~~~\nonumber\\
     (N_{14} c_\beta+ N_{13} s_\beta )+S_{i3}(N_{13}N_{14}  - \frac{\kappa}{\lambda} N_{15}^2 ) \big]
\end{eqnarray}

In the singlino-dominated-LSP scenario, taken $N_{11}=N_{12}=0$, the mass of LSP can be written as $m_{\tilde{\chi}^0_{\rm 1}} \approx M_{\tilde{\chi}^{0},\tilde{S} \tilde{S} } = 2 \kappa v_s$.
And from Eq.(\ref{Neu-mass}), Eq.(\ref{mass-sr}) and Eq.(\ref{mass-si}), one can find the sum rule \cite{sumrule}:
\begin{equation}
M_{\tilde{\chi}^{0},\tilde{S} \tilde{S} }^2  =  4 \kappa^2 v_s^2 = M_{S,S^R S^R}^2 + \frac{1}{3} M_{P,S^I S^I}^2 - \frac{4}{3} v_u v_d (\frac{\lambda^2 A_\lambda}{\mu}+\kappa)
\end{equation}
In the case that $h_1$ singlet-like, $\tan\beta$ sizable, $\lambda, \kappa$ and $A_\lambda$ not too large, this equation can become
\begin{equation}
    m_{\tilde{\chi}^0_{\rm 1}}^2 \approx m_{h_1}^2 + \frac{1}{3} m_{a_1}^2
\end{equation}

The chargino sector in the NMSSM is similar to neutralino sector. The charged higgsino $\tilde{H}_u^+$, $\tilde{H}_d^-$ (with mass scale $\mu$) and the charged gaugino $\tilde{W}^{\pm}$ (with mass scale $M_2$) can also mix respectively, forming two couples of physical chargino $\chi_1^{\pm}, \, \chi_2^{\pm}$.
In the gauge-eigenstate basis
$(\tilde{W}^+, \tilde{H}_{u}^+, \tilde{W}^-, \tilde{H}_{d}^-)$,
the chargino mass matrix is given by \cite{Ellwanger nmssm}
\begin{eqnarray}
M_{\tilde{C}}= \left( \begin{array}{cc} 0 & X^T \\ X & 0 \\ \end{array}
\right)\,,
\quad {\rm where }~
\label{chargino-mass}
X= \left( \begin{array}{cc}
    M_{2}                   & \sqrt{2} s_{\beta} m_W   \\
    \sqrt{2} c_{\beta} m_W  & \mu                      \\
\end{array} \right) \, .
\end{eqnarray}
To obtain the chargino mass eigenstates, one can use two unitary matrix to diagonalize the chargino mass matrix by
\begin{equation}
U^* X V^{-1} = M_{\tilde{\chi}^{\pm}}^{D}={\rm Diag}( m_{\tilde{\chi}^{\pm}_{\rm 1}},m_{\tilde{\chi}^{\pm}_{\rm 2}})
\end{equation}
where $M_{\tilde{\chi}^{\pm}}^{D}$ means the diagonal mass matrix, and the order of eigenvalues is $m_{\tilde{\chi}^{\pm}_{\rm 1}}<m_{\tilde{\chi}^{\pm}_{\rm 2}}$.
Meanwhile, we can get the mass eigenstates
\begin{equation}
\label{eq:chargino-rotation}
\left(
\begin{array}{c}
\tilde{\chi}^{+}_1 \\
\tilde{\chi}^{+}_2
\end{array} \right) = V_{ij} \left(
\begin{array}{c}
\tilde{W}^{+} \\
\tilde{H}^{+}_u
\end{array} \right)\,, \quad
\left(
\begin{array}{c}
\tilde{\chi}^{-}_1 \\
\tilde{\chi}^{-}_2
\end{array} \right) = U_{ij} \left(
\begin{array}{c}
\tilde{W}^{-} \\
\tilde{H}^{-}_d
\end{array} \right)\,.
\end{equation}

In the scNMSSM, since $M_2\gg \mu$, $\chi^\pm_1$ can be higgsino-dominated, with mass around $\mu$.
With $\chi^0_1$ singlino-dominated, $\chi^0_{2,3}$ can be higgsino-dominated, with masses nearly degenerate also around $\mu$, and with $N_{23}^2+N_{24}^2 > 0.5$.
Then with $\mu$ not large, smaller than other sparticle mass, the nearly-degenerate $\chi^\pm_1$ and $\chi^0_{2,3}$ can be called the next-to-lightest SUSY particles (NLSPs).
In this work, we will focus on the detection of the higgsino-dominated NLSPs ($\chi^\pm_1$ and $\chi^0_{2,3}$) in the scNMSSM.

\section{The Light Higgsino-dominated NLSPs in scNMSSM}
\label{sec:data}

In this work, we use the scan result in our former work on scNMSSM \cite{Wang-scNMSSM}, but only consider the surviving samples with singlino-dominated $\chi^0_1$ ($|N_{15}|^2>0.5$) as the LSP, and impose the SUSY search constraints with $\textsf{CheckMATE}$ \cite{checkmate2}.
We did the scan with the program $\textsf{NMSSMTools-5.4.1}$ \cite{nmssmtools}, and considered the constraints there, including theoretical constraints of vacuum stability and Landau pole, experimental constraints of Higgs data, muon g-2, B physics, dark matter relic density and direct searches, etc$^{2)}$.
\footnotetext{2) The detail of the constraints can be found in Ref.\cite{Wang-scNMSSM}. For the dark matter relic density, we only consider the upon bound, that is $\Omega h^2\leq 0.131$, considering there may be also other sources of dark matter.}
We also use $\textsf{HiggsBounds-5.1.1beta}$  \cite{higgsbounds} to constrain the Higgs sector (with $h_2$ as the SM-like Higgs and $123<m_{h_2}<127\GeV$), and $\textsf{SModelS-v1.1.1}$ \cite{smodels1} to constrain the SUSY particles.
The scanned spaces of the parameters are:
\begin{eqnarray}
\label{eq:space}
0<M_0<500\GeV, \,\, 0<M_{1/2}<2\TeV, \,\, |A_0|<10\TeV,
\nonumber\\
100<\mu<200\GeV,~~ \,\, 1<\tan\beta <30,~~~ \,\, 0.3<\lambda<0.7,
\nonumber\\
0<\kappa<0.7,\,\,~~~~~~~~|A_\lambda|<10\TeV,\,\,~~~~~ |A_\kappa|<10\TeV \, .
\end{eqnarray}

As shown in Ref.\cite{Wang-scNMSSM}, in the survived parameter space,
\begin{itemize}
  \item Glugino is heavier than 1.5 TeV, thus $M_{1/2}$ at GUT scale, or $M_3/2.4\simeq M_2/0.8 \simeq M_1/0.4$ at $M_{\rm SUSY}$ scale due to RGE runnings, is lager than about $700\GeV$.
  \item Due to the RGE runnings including $M_3$, the squarks can be heavy.
      Even the lightest squarks, e.g. $\tilde{t}_1$, are heavier than about $500\GeV$.
  \item Due to the RGE runnings including $M_2$ and $M_1$, the sleptons of the first two generations and all sneutrinos are heavier than about $300\GeV$, only the  $\tilde{\tau}_1$ can be lighter than that and to about $100\GeV$.
  \item The heavy charginos $\tilde{\chi}^\pm_2$ are wino-like and heavier than about $560\GeV$, the light charginos  $\tilde{\chi}^\pm_1$ are higgsino-like at $100\sim200 \GeV$.
  \item For the five neutralinos, $\tilde{\chi}^0_5$ is wino-like and heavier than $560\GeV$, the bino-dominated neutralino is heavier than $280\GeV$, the higgsino-dominated neutralinos are $100\sim200 \GeV$, while the singlino-dominated neutralino can be $60\sim400 \GeV$.
  \item For the Higgs sector, $h_2$ is the SM-like Higgs at $123\sim127 \GeV$, $h_1$ is singlet-dominated and lighter than $123\GeV$, the light CP-odd Higgs is singlet-dominated but can be lighter or heavier than $125\GeV$.
      Since $h_2$ is SM-like, the Higgs invisible decay caused by $m_{\tilde{\chi}^0_1}\simeq 60\GeV$ is at most about $20\%$, so as the Higgs exotic decays caused by $m_{h_1,a_1}\lesssim60\GeV$.
\end{itemize}
In this work, we choose the survived samples with $\tilde{\chi}^0_1$ LSP as singlino-dominated, and higgsino-dominated neutralino and chargino ($\tilde{\chi}^\pm_1$ and $\tilde{\chi}^0_{2,3}$) as NLSPs.
Thus the samples with higgsino-dominated neutralino as LSP, or $\tilde{\tau}_1$ as NLSP, are discarded.
In the following, we focus on the higgsino-dominated NLSPs in the scNMSSM, considering its constraints from direct search results at the LHC Run I and Run II, its production and decay, and its possibility of discovery at the HL-LHC in the future.

\subsection{Constraints from Direct SUSY Searches with CheckMATE}
In this work, we besides use \texttt{CheckMATE 2.0.26} \cite{checkmate2} to impose these constraints of direct SUSY search results at the LHC, which using all data at Run I and up to $36\fbm$ data at Run II \cite{cms039, ATLAS:2016uwq, CMS:2016zvj, checkmate2017}.
With masses at 100$\sim$200 GeV, the cross sections of the higgsino-dominated NLSPs can be sizeable, thus we pay special attention to the NLSPs.

Firstly, We use \texttt{MadGraph5\_aMC@NLO 2.6.6}\cite{madgraph} to generate three types of tree level processes at 8 TeV and 13 TeV:
\begin{equation}
\label{eq:process}
p p \to \tilde{\chi}^{+}_1 \tilde{\chi}^{-}_1 ~, ~~~
p p \to \tilde{\chi}^{\pm}_1 \tilde{\chi}^{0}_{2,3} ~,~~~
p p \to \tilde{\chi}^{0}_{2,3} \tilde{\chi}^{0}_{2,3} ~.
\end{equation}
Since the cross sections by the \texttt{MadGraph} are at tree level, we multiply them by a NLO K-factor calculated with the \texttt{Prospino2} \cite{prospino}.
Then, we use the \texttt{PYTHIA 8.2} \cite{pythia8} to deal with particle decay, parton showering, and hardronization, use \texttt{Delphes 3.4.1} \cite{delphe3} to simulate the detector response,
and use the anti-$k_T$ algorithm \cite{anti-kt} for jet clustering.

After the simulation, we can get a `.root' file. We use the \texttt{CheckMATE2} to read this `.root' file.
Then, we apply the same cuts in signal regions of the CMS and ATLAS experiments at 8 TeV and 13 TeV, by using analysis cards which have been implemented in \texttt{CheckMATE2}.
At the last step, with the \texttt{CheckMATE2} we get a r-value for each samples, which is defined as
\begin{equation}
\label{r-value}
r \equiv \frac{S - 1.64\Delta S}{S_{Exp.}^{95}}
\end{equation}
where $S$ is the total number of expected signal events, $\Delta S$ is the uncertainty of $S$, and $S_{Exp.}^{95}$ is the experimentally measured 95\% confidence limit of signal events number.
So, a model can be considered excluded at 95\% confidence level, if $r \geq 1$.
If the $r \geq 1$ in only one signal region, the model can also be excluded.
We can get $r_{max}$, the maximal value of $r$ in different signal regions.
The model is excluded if $r_{max} \geq 1$.

After using \texttt{CheckMATE} to checking our surviving samples, we notice that most of the samples excluded are by the CMS analysis in multilepton final states at 13 TeV LHC with $35.9 \fbm$ data \cite{cms039}$^{3)}$.
\footnotetext{3) The CMS collaboration has not released update results with more data in the multilepton channel up to now. The cut scheme of ATLAS analysis in this channel is different, and we checked that the ATLAS result with $13.3\fbm$ data \cite{ATLAS:2016uwq} implemented in the $\textsf{CheckMATE}$ is much weaker than the CMS result in constraining our samples, and thus even with $139\fbm$ data \cite{ex-susy-3l} it has no significant impact on our final conclusion. We also have considered the CMS analysis for the compressed spectrum with $12.9\fbm$ data \cite{CMS:2016zvj}, and find that even with the current $139\fbm$ data it can not constrain our samples more.}
We checked that the relevant mechanism is $\tilde{\chi}_1^{\pm} \tilde{\chi}_2^{0}$ produced and each decaying to 2 bodies.
Since the sleptons are heavier, the $\tilde{\chi}_1^{\pm}$ and $\tilde{\chi}_2^{0}$ mainly decay to the $\tilde{\chi}_1^{0}$ LSP plus a $W$, or $Z$, or Higgs boson.
The most effective processes excluding the samples are $pp\to \tilde{\chi}_1^{\pm}(W^{\pm}\tilde{\chi}_1^{0}) \tilde{\chi}_2^{0} (Z\tilde{\chi}_1^{0})$ and $pp\to \tilde{\chi}_1^{\pm}(W^{\pm}\tilde{\chi}_1^{0}) \tilde{\chi}_2^{0} (h_2\tilde{\chi}_1^{0})$.

The searching strategy for these two processes is three or more leptons plus large $\pslash_T$ in the final state, since these channels are relatively cleaner than the jet channels at the LHC.
The CMS searches related to our processes included the following signal regions (SR) SR-A, SR-C and SR-F
\begin{itemize}
    \item \textbf{SR-A:} events with three light leptons (e or $\mu$), two of which forming an opposite sign same-flavor (OSSF) pair. The SR-A is divided into 44 bins, according to the invariant mass of OSSF pair $M_{\ell \ell}$, the third lepton's transverse mass $M_T$ and the missing transverse momentum $\pslash_T$.
    The transverse mass $M_T$ is defined as
    \begin{eqnarray}
    M_T = \sqrt{2 p_T^{\ell} \SLASH{p}{.2}_{T}[1-\cos(\Delta\phi)]} \;,
    \end{eqnarray}
    where $\Delta\phi$ is the angle between ${\vec p}_T^{\ell}$ and ${\SLASH{\vec p}{.2}_{T}}$.
    \item \textbf{SR-C:} events with two light leptons (e or $\mu$) forming an OSSF pair, and one $\uptau_{\rm h}$ candidate.
        The SR-C is divided into 18 bins, according to the invariant mass $M_{\ell \ell}$, the two-lepton `stransverse mass' $M_{T2}(\ell_1, \ell_2)$ \cite{mt2} instead of $M_T$ on the off-Z regions, and the $\pslash_T$.
        The $M_{T2}$ is defined as
    \begin{equation}
        M_{T2}=  \min \limits_{\pslash_{T}^1+\pslash_{T}^2=\pslash_T} \left [ \max \left \{ M_T( {\vec p}_{T}^{\ell_1}, \pslash_{T}^1), M_T({\vec p}_{T}^{\ell_2}, \pslash_{T}^2) \right \} \right] ~,
    \end{equation}
    where $\vec{p}_T^{\ell_1}$ and $\vec{p}_T^{\ell_2}$ are the transverse momentum for the two leptons respectively, while $\pslash_{T}^1$ and $\pslash_{T}^2$ stand for the random two components of the missing transverse momentum $\pslash_T$.
    And it is used to suppressed the SM background, since the large $t\bar{t}$ background is at low $M_{T2}$.
    \item \textbf{SR-F:} events with one light lepton (e or $\mu$) plus two $\uptau_{\rm h}$ candidates. SR-F is divided into 12 bins, according to $M_{\ell \ell}$, $M_{T2}(\ell, \uptau_{1})$ and the $\pslash_T$.
\end{itemize}

Recently after 2017, ATLAS and CMS collaborations have released several new search results with more Run-II data up to $139\fbm$ in such channels as $2\ell+\pslash_T$ \cite{ex-susy-2l}, $3\ell+\pslash_T$ \cite{ex-susy-3l} (see also non-SUSY interpretations of multilepton anomalies in Refs.\cite{multilepton-exotic}), $2\gamma+\pslash_T$ \cite{ex-susy-2photon}, Higgs$+\pslash_T$ \cite{ex-susy-higgs}, and Higgs$+\ell+\pslash_T$ \cite{susy-lbb}, and for compressed mass spectrum \cite{susy-compressed}.
In their analyses, they considered simple models, where purely higgsino or wino NLSP produced in pair, each decaying to $\tilde{\chi}^0_1$ plus $h$, $Z$, or $W^\pm$ in $100\%$, so the results do not apply to our samples directly.
Imposing these new constraints is fussy, so we tend to put them in our later works.

\subsection{Production and decay of the higgsino-dominated NLSPs}

For the surviving samples we first checked the production cross sections of
$\tilde{\chi}_1^{+} \tilde{\chi}_2^{0}$, $\tilde{\chi}_1^{+} \tilde{\chi}_3^{0}$,
$\tilde{\chi}_1^{-} \tilde{\chi}_2^{0}$, $\tilde{\chi}_1^{-} \tilde{\chi}_3^{0}$,
$\tilde{\chi}_1^{+} \tilde{\chi}_1^{-}$, $\tilde{\chi}_2^{0} \tilde{\chi}_3^{0}$,
$\tilde{\chi}_2^{0} \tilde{\chi}_2^{0}$ and $\tilde{\chi}_3^{0} \tilde{\chi}_3^{0}$ at the 14 TeV LHC.
Since the NLSPs are higgsino-dominated, the cross sections are not much different from those of pure higgsino production.
Here we only revise some of the old conclusions.
\begin{itemize}
\item The cross sections reduce quickly when the masses increasing, for the partonic Mandelstam variable $\hat{s}$ increase, and the parton fluxes decrease.
\item The cross section of $\tilde{\chi}_1^{+} \tilde{\chi}_i^{0}$ is about 2 times of $\tilde{\chi}_1^{-} \tilde{\chi}_i^{0}$, for both $i=2$ and $3$.
    The reason is that the LHC is proton-proton collider, and the parton distribution functions (PDF) for up quark is larger than down quark.
\item The cross sections of $\tilde{\chi}_2^{0} \tilde{\chi}_2^{0}$ and $\tilde{\chi}_3^{0} \tilde{\chi}_3^{0}$ are very small, only a few $\rm fb$.
The reason is that the squarks are very heavy, so $pp \to \tilde{\chi}_i^{0} \tilde{\chi}_i^{0}$ are mainly produced from s channel through $Z$ boson resonance.
The coupling of $Z-\tilde{\chi}_i^{0} - \tilde{\chi}_j^{0}$ is given by
\begin{align}
\label{coupling-zxx}
C_{Z \tilde{\chi}_i^{0}  \tilde{\chi}_j^{0} } =
& -\frac{i}{2} (g_1 s_W   + g_2 c_W  )(N^*_{j 3} N_{{i 3}}  - N^*_{j 4} N_{{i 4}} )(\gamma_{\mu} P_L)      \notag  \\
& + \,\frac{i}{2} (g_1 s_W   + g_2 c_W  )(N^*_{i 3} N_{{j 3}}  - N^*_{i 4} N_{{j 4}} )(\gamma_{\mu}\ P_R)
\end{align}
where the matrix $N$ is neutralino mixing matrix.
Since $\tilde{\chi}_{2,3}^0$ are higgsino-dominated, $m_{\tilde{\chi}_{2,3}^0}\simeq \mu$, thus from Eq.(2.30) we know $N_{i3}:N_{i4} \approx -1$,
then $(|N_{i3}|^2-|N_{i4}|^2) \approx 0 $, and $\sigma(pp \to \tilde{\chi}_i^{0}  \tilde{\chi}_i^{0}) \approx 0$.
\end{itemize}

\begin{figure*}[!htb]
  \centering
\includegraphics[width=16cm]{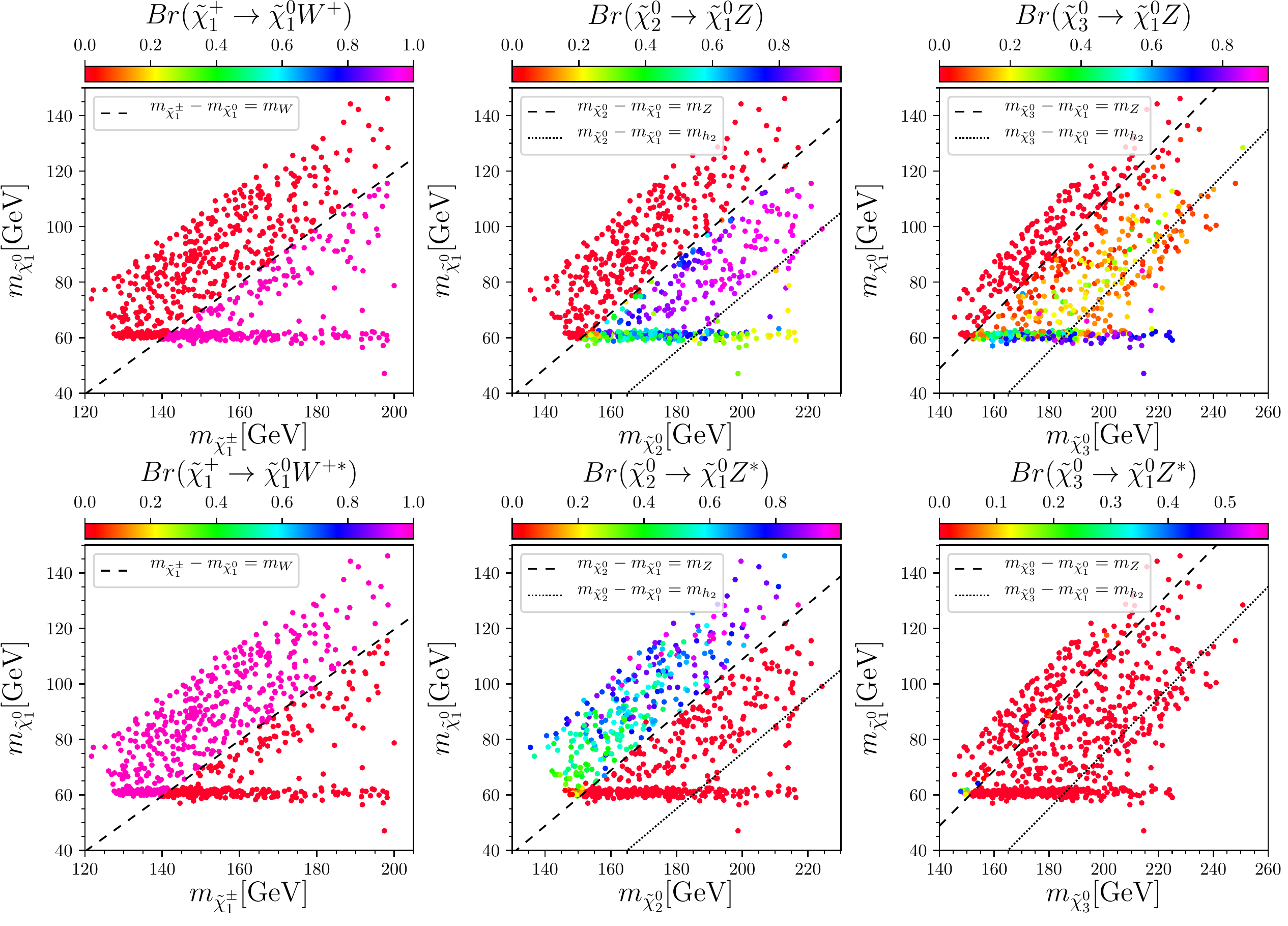}
\figcaption{\label{fig1}(color online) The samples in the $m_{\tilde{\chi}_1^{0}}$ versus $m_{\tilde{\chi}_i^{0}}$ planes (left $i=\pm$, middle $i=2$ and right $i=3$).
The colors indicate the branching ratios of the chargino $\tilde{\chi}_1^{+}$ to $\tilde{\chi}_1^{0}$ plus $W$ boson, and neutralino $\tilde{\chi}^0_{2,3}$ to $\tilde{\chi}_1^{0}$ plus $Z$ boson respectively.
In the upper panel, the $W/Z$ boson is a real one and the decay is real two-body decay; while in the lower panel the $W/Z$ boson is a virtual one and the decay is virtual three-body decay.}
\end{figure*}

\begin{figure*}[!htb]
  \centering
\includegraphics[width=16cm]{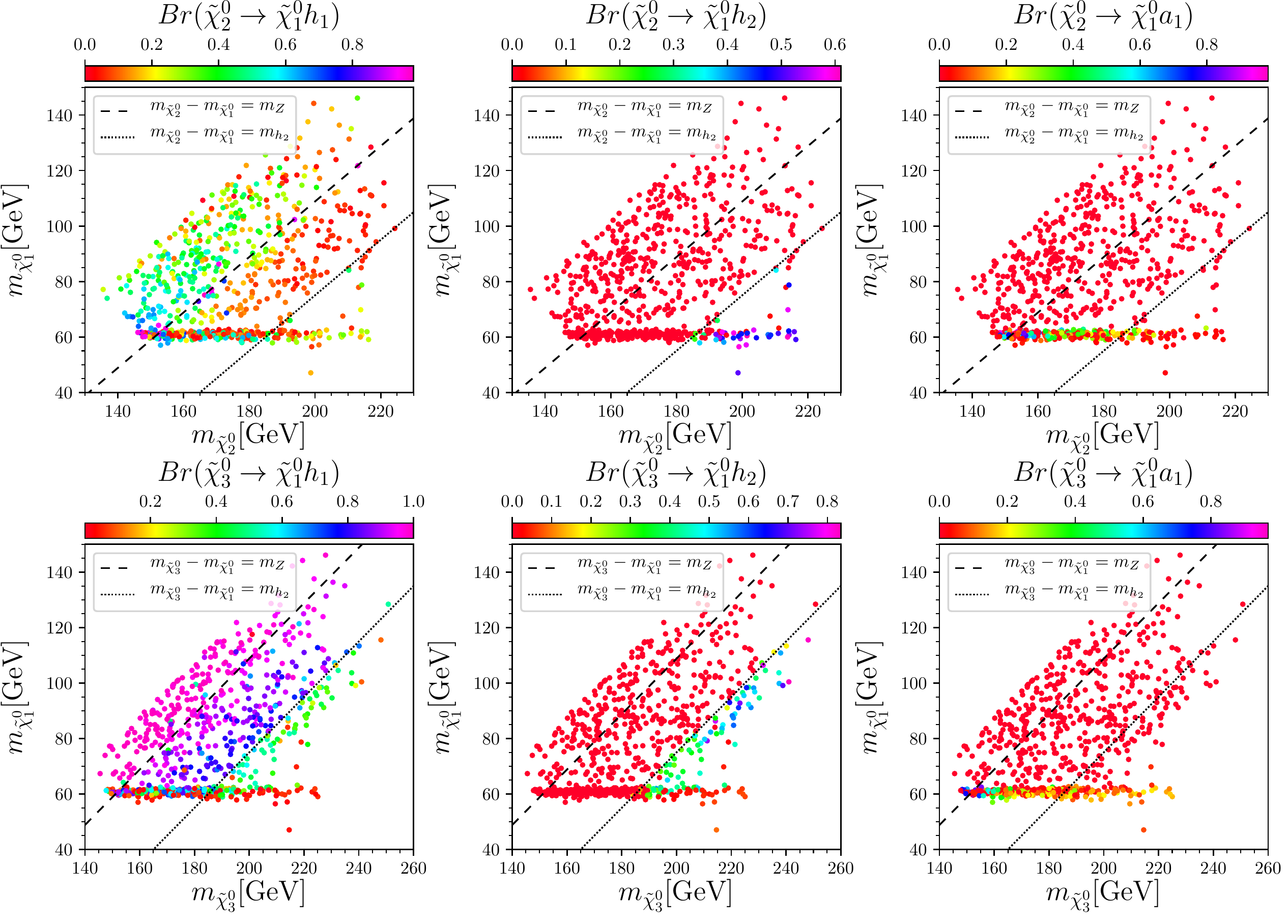}
\figcaption{\label{fig2}(color online) The samples in the $m_{\tilde{\chi}_1^{0}}$ versus $m_{\tilde{\chi}_i^{0}}$ planes (upper $i=2$, lower $i=3$).
From left to the right, colors indicate the branching ratios
$Br(\tilde{\chi}_i^{0} \to \tilde{\chi}_1^{0} h_1)$,
$Br(\tilde{\chi}_i^{0} \to \tilde{\chi}_1^{0} h_2)$, and
$Br(\tilde{\chi}_i^{0} \to \tilde{\chi}_1^{0} a_1)$, respectively.
The dashed line and the dotted line means that the mass difference, $m_{\tilde{\chi}_i^0}-m_{\tilde{\chi}_1^{0}}$, equal to $m_Z$ and $m_{h_2}$ respectively.}
\end{figure*}

The decay branching ratios of the NLSPs are shown in Fig.\ref{fig1} and Fig.\ref{fig2}.
From the left plots in Fig.\ref{fig1}, we can see that, the chargino $\tilde{\chi}_1^{\pm}$ decay to $\tilde{\chi}_1^{0}$ plus a W boson in $100\%$: when the mass difference between $\tilde{\chi}_1^{\pm}$ and $\tilde{\chi}_1^{0}$ is larger than $m_W$, the W boson is a real one; while when the mass difference is insufficient, the W boson is a virtual one, or the decay is a three-body decay.  
The low mass difference is negative for us to search for the SUSY particles, since the leptons coming form a virtual W boson are very soft and hard to detect.

The main decay modes of neutralino $\tilde{\chi}_i^{0}$ ($i=2,3$) are to a $\tilde{\chi}_1^{0}$ plus a Z boson or a Higgs boson.
In the middle and right plots of Fig.\ref{fig1} and in Fig.\ref{fig2}, we show the branching ratios of the neutralinos $\tilde{\chi}_i^{0}$ on the plane of $m_{\tilde{\chi}_1^{0}}$ vs $m_{\tilde{\chi}_i^{0}}$, where $i=2,3$.
In these plots, we use the dashed line, $m_{\tilde{\chi}_i^0}-m_{\tilde{\chi}_1^{0}}=m_Z$, and the dotted line, $m_{\tilde{\chi}_i^0}-m_{\tilde{\chi}_1^{0}}=m_{h_2}$, dividing the plane into 3 parts.
\begin{itemize}
    \item \textbf{Case I:} In the region $m_{\tilde{\chi}_i^0}-m_{\tilde{\chi}_1^{0}} < m_Z$, the neutralino $\tilde{\chi}_i^{0}$ can only decay to $\tilde{\chi}_1^{0}$ plus a virtual $Z$ boson or a light Higgs boson $h_1$.
    And we can see from the lower middle plot of Fig.\ref{fig1} and the upper left plot of Fig.\ref{fig2}, the $\tilde{\chi}_2^{0}$ mainly decay to a virtual Z boson plus a $\tilde{\chi}_1^{0}$, with only a small fraction to the light Higgs boson $h_1$ plus $\tilde{\chi}_1^{0}$.
    On the contrary, the lower left plot of Fig.\ref{fig2} shows that the $\tilde{\chi}_3^{0}$ mainly decay to the light Higgs boson $h_1$.

    \item \textbf{Case II:} In the region $m_Z \le m_{\tilde{\chi}_i^0}-m_{\tilde{\chi}_1^{0}} < m_{h_2}$, the neutralino $\tilde{\chi}_i^{0}$ can decay to $\tilde{\chi}_1^{0}$ plus a real Z boson or a light Higgs boson $h_1/a_1$.
        As showed in the upper middle plot of Fig.\ref{fig1}, the $\tilde{\chi}_2^{0}$ mainly decay to a real Z boson plus $\tilde{\chi}_1^{0}$.
        While according to the upper right plot of Fig.\ref{fig1} and lower left plot of Fig.\ref{fig2}, the $\tilde{\chi}_3^{0}$ mainly decay to a light Higgs boson $h_1$ plus $\tilde{\chi}_1^{0}$.

    \item \textbf{Case III:} In the region $m_{\tilde{\chi}_i^0}-m_{\tilde{\chi}_1^{0}} \ge m_{h_2}$, all these decay channels are opened.
        The $\tilde{\chi}_2^{0}$ mainly decay to $\tilde{\chi}_1^{0}$ plus a 125 GeV SM-like Higgs boson $h_2$, while $\tilde{\chi}_3^{0}$ mainly decay to $\tilde{\chi}_1^{0}$ plus $Z$ bosons.
\end{itemize}
In the channel $\tilde{\chi}_i^{0} \to \tilde{\chi}_1^{0} Z$ ($i=2,3$), like the $\tilde{\chi}_1^{\pm}\to W^\pm \tilde{\chi}_1^0$, when the mass difference is insufficient the Z boson also becomes a virtual one.
In the channel $\tilde{\chi}_i^{0} \to \tilde{\chi}_1^{0} H$ ($i=2,3$), where the Higgs boson can be $h_1$ or $h_2$, and $h_2$ is the SM-like one.
Both $h_1$ and $h_2$ mainly decay to $b \bar{b}$, thus the $t\bar{t}$ background is sizable at the LHC.
In the case that Higgs decay to WW, ZZ, or $\uptau\uptau$, and W or Z decays leptonically, it might contribute to the multilepton final state.
Since the light Higgs $h_1$ is highly singlet-dominated, the $\tilde{\chi}_i^{0} \to \tilde{\chi}_1^{0} h_1$ is very hard to contribute to the multilepton signal regions.
Thus only the $\tilde{\chi}_i^{0} \to \tilde{\chi}_1^{0} h_2$ can contribute to the multilepton signal regions visibly.

It is worth to mention that, when the heavier neutralinos decay to the $\tilde{\chi}_1^{0}$ LSP, the $\tilde{\chi}_2^{0}$ and $\tilde{\chi}_3^{0}$ behave differently.
Especially in the case II,
$\tilde{\chi}_2^{0}$ prefers to decay to a Z boson plus $\tilde{\chi}_1^{0}$, $Br(\tilde{\chi}_2^{0} \to \tilde{\chi}_1^{0} Z) > Br(\tilde{\chi}_2^{0} \to \tilde{\chi}_1^{0} h_1)$;
while $\tilde{\chi}_3^{0}$ tends to decay to a light Higgs boson $h_1$ plus $\tilde{\chi}_1^{0}$, $Br(\tilde{\chi}_3^{0} \to \tilde{\chi}_1^{0} Z) < Br(\tilde{\chi}_3^{0} \to \tilde{\chi}_1^{0} h_1)$.
The couplings $C_{h_1 \tilde{\chi}_2^{0} \tilde{\chi}_1^{0}}$ and $C_{h_1 \tilde{\chi}_3^{0} \tilde{\chi}_1^{0}}$ can be written down as
\begin{equation}
C_{h_1 \tilde{\chi}_2^{0} \tilde{\chi}_1^{0} } \sim \frac{\lambda  \left(N_{1 4} N_{2 3}+N_{1 3} N_{2 4}\right)
   S_{1 3}}{\sqrt{2}}-\sqrt{2} \kappa  N_{1 5} N_{2 5}
   S_{1 3}
\end{equation}
\begin{equation}
C_{h_1 \tilde{\chi}_3^{0} \tilde{\chi}_1^{0} } \sim \frac{\lambda  \left(N_{1 4} N_{3 3}+N_{1 3} N_{3 4}\right)
   S_{1 3}}{\sqrt{2}}-\sqrt{2} \kappa  N_{1 5} N_{3 5}
   S_{1 3}
\end{equation}
where the $N_{1 1}$, $N_{1 2}$, $N_{2 1}$, $N_{2 2}$, $N_{3 1}$ and $N_{3 2}$ was set to $0$ since the wino and bino are very heavy and decoupled in the scNMSSM, and the $S_{1 1}$ and $S_{1 2}$ was set to $0$ since $|S_{1 3}|\gg |S_{1 1}|, |S_{1 2}|$.
$\lambda / \sqrt{2} \ll 1$ and $\sqrt{2} \kappa \ll 1$, so the couplings $C_{h_1 \tilde{\chi}_2^{0} \tilde{\chi}_1^{0}}$ and $C_{h_1 \tilde{\chi}_3^{0} \tilde{\chi}_1^{0}}$ are both very small and roughly the same.
While the couplings $C_{Z \tilde{\chi}_2^{0} \tilde{\chi}_1^{0}}$ and $C_{Z \tilde{\chi}_3^{0} \tilde{\chi}_1^{0}}$ can be different from each other according to Eq.(\ref{coupling-zxx}), which can be approximated to
\begin{equation}
\label{z21}
C_{Z \tilde{\chi}_2^{0} \tilde{\chi}_1^{0} } \sim \frac{g_2}{c_W} (N_{1 3} N_{2 3}  - N_{1 4} N_{2 4})
\end{equation}
\begin{equation}
\label{z31}
C_{Z \tilde{\chi}_3^{0} \tilde{\chi}_1^{0} } \sim \frac{g_2}{c_W} (N_{1 3} N_{3 3}  - N_{1 4} N_{3 4})
\end{equation}
where the $g_2 / c_W \sim 1$.
When the two terms in Eq.(\ref{z21}) or Eq.(\ref{z31}) have different sign, and do not cancel with each other, the couplings $C_{Z \tilde{\chi}_i^{0} \tilde{\chi}_1^{0}}$ can be much larger than $C_{h_1 \tilde{\chi}_i^{0} \tilde{\chi}_1^{0}}$;
otherwise the cancel between the two terms can make $C_{Z \tilde{\chi}_i^{0} \tilde{\chi}_1^{0}}$ smaller than $C_{h_1 \tilde{\chi}_i^{0} \tilde{\chi}_1^{0}}$.
For some surviving samples, $C_{Z \tilde{\chi}_3^{0} \tilde{\chi}_1^{0}}$ have the cancellation between the two terms,
and that leads to small $Br(\tilde{\chi}_3^{0} \to \tilde{\chi}_1^{0} Z)$ and large  $Br(\tilde{\chi}_3^{0} \to \tilde{\chi}_1^{0} h_1)$.
Six benchmark points are listed in the Table \ref{tab:benchmark}.

\begin{table*}[!htb]
\centering
\tabcaption{\label{tab:benchmark}Masses and branching ratios for 6 benchmark points in the scNMSSM.
The signal significances in the last line are calculated with luminosity of $300\fbm$, and with similar analysis of multi-lepton final state as in Ref. \cite{cms039}.}
\footnotesize
\vspace{-1mm}
\begin{tabular*}{120mm}{|c@{\extracolsep{\fill}}|c|c|c|c|c|c|}
\toprule
\hline
       & P1    & P2    & P3    & P4    & P5    & P6    \\ \hline
$m_{\tilde{\chi}_1^{\pm}}$ (GeV)  & 183   & 173   & 175   & 189   & 175   & 187   \\ \hline
$m_{\tilde{\chi}_1^{0}}$ (GeV)  & 120   & 119   & 103   & 108   & 82    & 92    \\ \hline
$m_{\tilde{\chi}_2^{0}}$ (GeV)  & 200  & 187  & 200  & 209  & 202  & 206  \\ \hline
$m_{\tilde{\chi}_3^{0}}$ (GeV)  & 216   & 200   & 214   & 216   & 212   & 209   \\ \hline
$Br(\tilde{\chi}_1^{+} \to \tilde{\chi}_1^{0} W)$  & 0\%   & 0\%   & 0\%   & 100\% & 100\% & 100\% \\ \hline
$Br(\tilde{\chi}_1^{+} \to \tilde{\chi}_1^{0} W^*)$ & 100\% & 100\% & 100\% & 0\%   & 0\%   & 0\%   \\ \hline
$Br(\tilde{\chi}_2^{0} \to \tilde{\chi}_1^{0} Z)$  & 0\%   & 0\%   & 90\%  & 93\%  & 94\%  & 95\%  \\ \hline
$Br(\tilde{\chi}_2^{0} \to \tilde{\chi}_1^{0} Z^*)$ & 80\%  & 65\%  & 0\%   & 0\%   & 0\%   & 0\%   \\ \hline
$Br(\tilde{\chi}_2^{0} \to \tilde{\chi}_1^{0} h_1)$ & 20\%  & 35\%  & 10\%  & 7\%   & 6\%   & 5\%   \\ \hline
$Br(\tilde{\chi}_2^{0} \to \tilde{\chi}_1^{0} h_2)$ & 0\%   & 0\%   & 0\%   & 0\%   & 0\%   & 0\%   \\ \hline
$Br(\tilde{\chi}_3^{0} \to \tilde{\chi}_1^{0} Z)$  & 1\%   & 0\%   & 1\%   & 13\%  & 13\%  & 38\%  \\ \hline
$Br(\tilde{\chi}_3^{0} \to \tilde{\chi}_1^{0} Z^*)$ & 0\%   & 0\%   & 0\%   & 0\%   & 0\%   & 0\%   \\ \hline
$Br(\tilde{\chi}_3^{0} \to \tilde{\chi}_1^{0} h_1)$ & 99\%  & 100\% & 99\%  & 87\%  & 33\%  & 62\%  \\ \hline
$Br(\tilde{\chi}_3^{0} \to \tilde{\chi}_1^{0} h_2)$ & 0\%   & 0\%   & 0\%   & 0\%   & 54\%  & 0\%   \\ \hline
$ss=S/\sqrt{B} @ 300 \rm fb^{-1}$ $(\sigma)$  & 3.1   & 2.9   & 2.1   & 3.8   & 10.8  & 8.8   \\ \hline
\bottomrule
\end{tabular*}
\end{table*}

\subsection{Possibility of discovery at the HL-LHC in the future}

In this part, we investigate the possibility of detect electroweakinos at the future High Luminosity LHC (HL-LHC).
We adopt the same analysis of multilepton final state by CMS \cite{cms039}, only increasing the integrated luminosity from 35.9 fb$^{-1}$ to 300 fb$^{-1}$, to see the possibility of discovery in the future.
And we evaluate the signal significance by
\begin{equation}
    ss=S/\sqrt{B}
\end{equation}
where $S$ and $B$ are the number of events from signal and background processes respectively.

\begin{figure*}[!htb]
  \centering
\includegraphics[width=16cm]{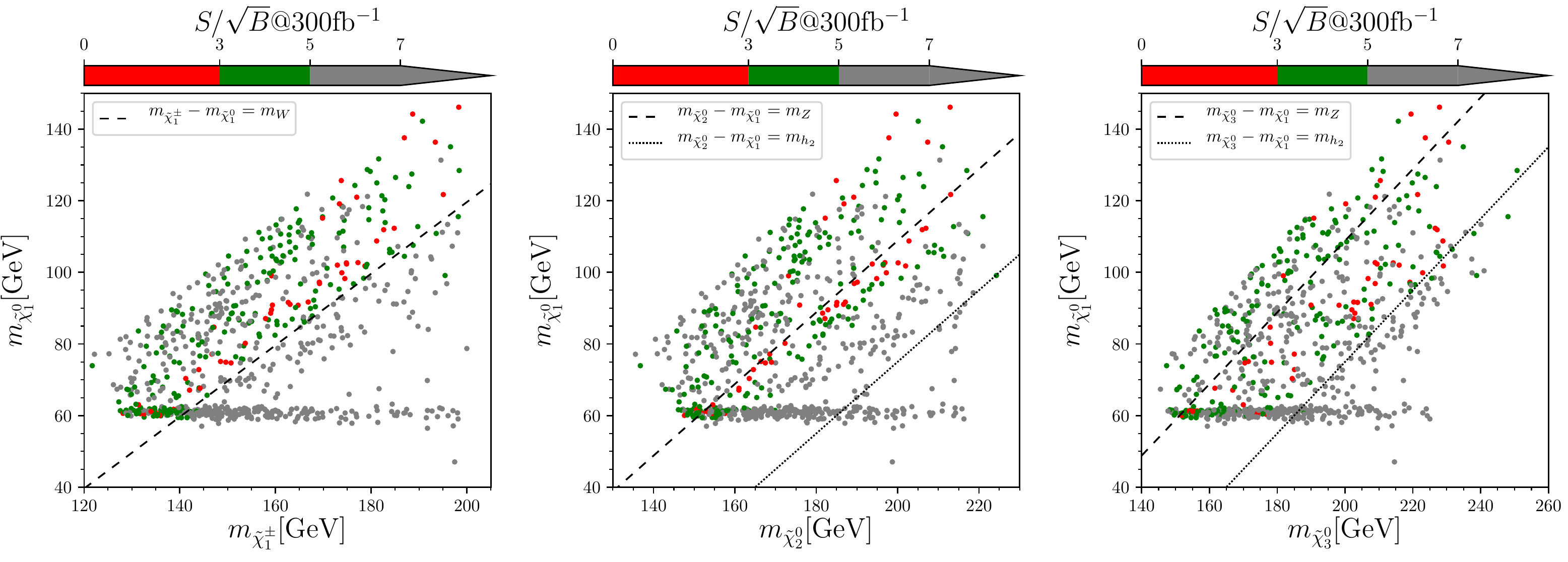}
\figcaption{\label{fig3}(color online) The samples in the $m_{\tilde{\chi}_1^{0}}$ versus $m_{\tilde{\chi}_1^{\pm}}$ (left), $m_{\tilde{\chi}_1^{0}}$ versus $m_{\tilde{\chi}_2^{0}}$ (middle), $m_{\tilde{\chi}_1^{0}}$ versus $m_{\tilde{\chi}_3^{0}}$ (right) planes.
The colors indicates the signal significance, where red represents $ss<3\sigma$, green represents $3\sigma<ss<5\sigma$, and gray represents $ss > 5\sigma$.
In the left plane, the dashed line indicates the mass difference equal to $m_W$, $m_{\tilde{\chi}_1^{\pm}}-m_{\tilde{\chi}_1^{0}}=m_W$.
In the middle and right planes, the dashed line and dotted line indicate the mass difference equal to $m_Z$ and $m_{h_2}$ respectively, that is, $m_{\tilde{\chi}_i^{0}}-m_{\tilde{\chi}_1^{0}}=m_Z$ and $m_{\tilde{\chi}_i^{0}}-m_{\tilde{\chi}_1^{0}}=m_{h_2}$, where $i=2,3$ for the middle and right planes respectively.}
\end{figure*}

\begin{figure*}[!htb]
\centering
\includegraphics[width=6cm]{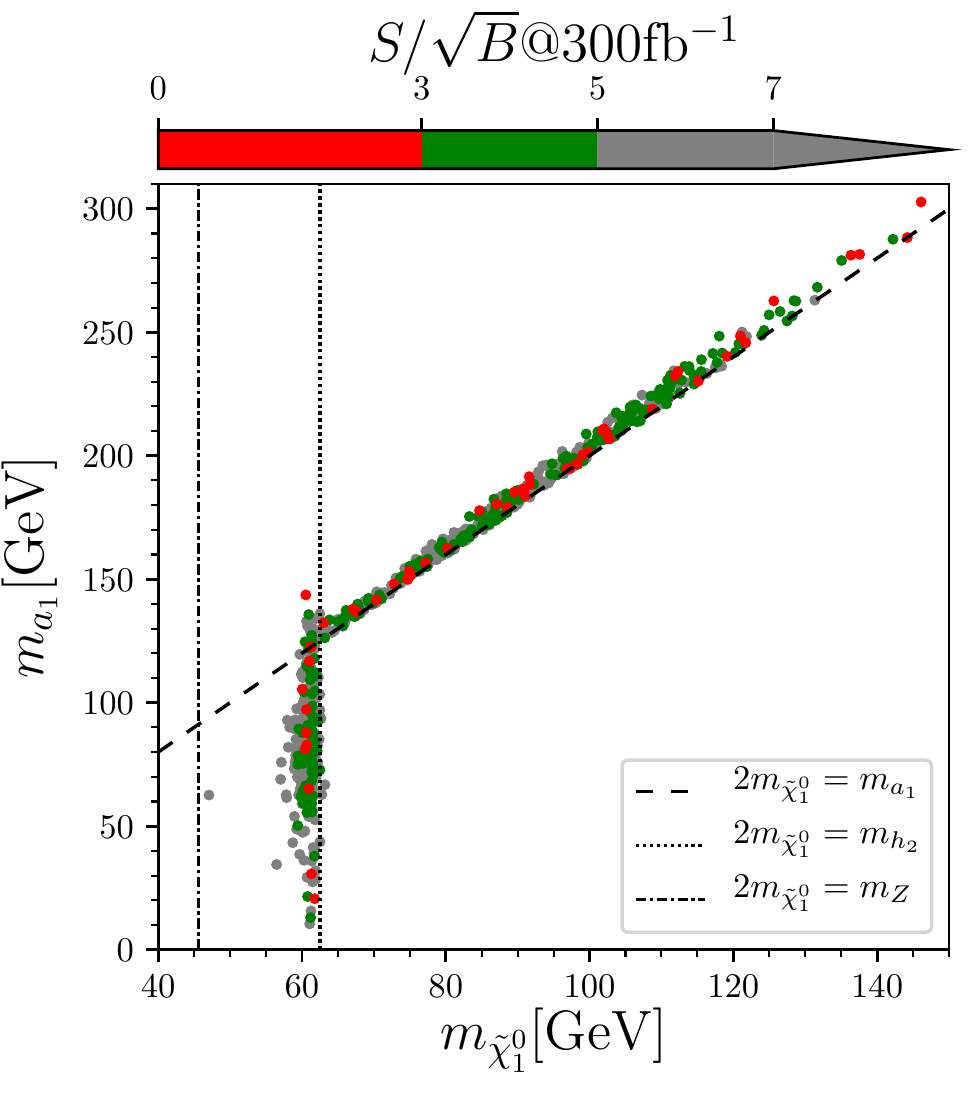}
\figcaption{\label{fig4}(color online) The samples in the $m_{a_1}$ versus $m_{\tilde{\chi}_1^{0}}$ plane.
The color convention is the same as in Fig.\ref{fig3}.
The dashed, dotted and dash-dotted lines indicate $2m_{\tilde{\chi}_1^{0}}$, equal to $m_{a_1}$, $m_{h_2}$, and $m_Z$ respectively.}
\end{figure*}

In Fig.\ref{fig3}, we show $ss$ on the planes of $m_{\tilde{\chi}_1^{0}}$ versus $m_{\tilde{\chi}_1^{\pm}}$, $m_{\tilde{\chi}_2^{0}}$ and $m_{\tilde{\chi}_3^{0}}$ respectively.
We can see that most of the samples can be checked at $5\sigma$ level when the mass difference between LSP $\tilde{\chi}_1^{0}$ and NLSPs $\tilde{\chi}_1^{\pm}, \tilde{\chi}_{2,3}^{0}$ is sufficient.
However, there are still some samples that can not be checked at level above 3 or 5 sigma.
The main reason is that the mass spectra is compressed, so that the leptons from the decay of NLSPs are very soft.
Because $\pslash_T$ cut has to be very large at the LHC due to the large background, detecting soft particles is not easy.
Combining with Fig.\ref{fig1} and \ref{fig2}, we can learn the following facts:
\begin{itemize}
    \item If $\tilde{\chi}_1^{\pm}$ or $\tilde{\chi}_i^{0}$ ($i=2,3$) decays to a virtual vector boson, or in the area upon the dashed line in all the plots, the signal significance is less than $5\sigma$, and it is hard to check with $300\fbm$ data at the LHC.
    \item If $\tilde{\chi}_1^{\pm}$ or $\tilde{\chi}_2^{0}$ decays to a real vector boson, the area between the dashed and dotted line in left and middle plots, the signal significance can be larger than $5\sigma$, and it is easy to check with $300\fbm$ data at the LHC.
    \item The $\tilde{\chi}_3^{0}$ decay to a light Higgs $h_1$, the area between the dashed and dotted line in the right plane, the signal significance is less than $5\sigma$ for some samples.
        The reason is that the light Higgs $h_1$ mainly decay to $b\bar{b}$, which is hard to distinguish from the background.
    \item If $\tilde{\chi}_i^{0}$ ($i=2,3$) decays to an SM-like Higgs, or in the area below the dotted line in middle and right plots, the signal significance is also larger than $5\sigma$ for most samples.
\end{itemize}
For the samples with insufficient mass difference between the NLSPs and LSP, the integrate luminosity 300 fb$^{-1}$ is not enough.
So we also tried to increase the luminosity to 3000 fb$^{-1}$, the result is that nearly all samples can be checked with $ss > 5$ at 3000 fb$^{-1}$.

In Fig.\ref{fig4}, we show the signal significance $ss$ on the planes of $m_{\tilde{\chi}_1^{0}}$ versus $m_{a_1}$.
We can see that there are mainly two mechanisms for dark matter annihilation:
the $a_1$ funnel where $2m_{\tilde{\chi}_1^{0}}\simeq m_{a_1}$, and the  $h_2/Z$ funnel where $2m_{\tilde{\chi}_1^{0}}\simeq m_{h_2}$ or $2m_{\tilde{\chi}_1^{0}}\simeq m_{Z}$.
Unfortunately, searching for the NLSPs is helpless in distinguishing these two mechanisms.
While as shown in Ref.\cite{Wang-scNMSSM}, the spin-independent cross section of $\tilde{\chi}^0_1$ can be sizable, so such singlino-dominated dark matter may be accessible in the future direct detections, such as XENONnT and LUX-ZEPLIN (LZ-7 2T).

\section{Conclusions}
\label{conclusion}

In this work, we have discussed the light higgsino-dominated NLSPs in the scNMSSM, which is also called the non-universal Higgs mass (NUHM) version of the NMSSM.
We use the scenario with singlino-dominated $\tilde{\chi}^0_1$ and SM-like $h_2$ in the scan result in our former work on scNMSSM, where we considered the constraints including theoretical constraints of vacuum stability and Landau pole, experimental constraints of Higgs data, muon g-2, B physics, dark matter relic density and direct searches, etc.
In our scenario, the bino and wino are heavy because of the high mass bound of gluino and the unification of gaugino masses at the GUT scale.
Thus the $\tilde{\chi}^{\pm}_1$ and $\tilde{\chi}^0_{2,3}$ are higgsino-dominated and mass-degenerated NLSPs.

We first investigate the direct constraints to these light higgsino-dominated NLSPs, including searching for SUSY particle at the LHC Run-I and Run-II.
We use Monte Carlo to do detailed simulations to impose these constraints from search SUSY particles at the LHC.
Then we discuss the possibility of checking the light higgsino-dominated NLSPs at the HL-LHC in the future.
We use the same analysis by increasing the integrated luminosity to 300 fb$^{-1}$ and 3000 fb$^{-1}$.

Finally, we come to the following conclusions regarding the higgsino-dominated $100\sim200\GeV$ NLSPs in scNMSSM:
\begin{itemize}
    \item Among the search results for electroweakinos, the `multi-lepton final state' constrain our scenario most, and can exclude some of our samples. While with all data at Run I and up to $36\fbm$ data at Run II at the LHC, the search results by ATLAS and CMS can still not exclude the light higgsino-dominated NLSPs of $100\sim200\GeV$.
    \item When the mass difference with $\tilde{\chi}^0_{1}$ is smaller than $m_{h_2}$, $\tilde{\chi}^0_{2}$ and $\tilde{\chi}^0_{3}$ have different preference on decaying to $Z/Z^*$ or $h_1$.
    \item The best channels to detect the NLSPs are though the real two-body decay $\tilde{\chi}^{\pm}_{1} \to \tilde{\chi}^0_1 W$ and $\tilde{\chi}^0_{2,3} \to \tilde{\chi}^0_1 Z/h_2$.
        When the mass difference is sufficient, most of the samples can be checked at 5 $\sigma$ level with future $300\fbm$ data at the LHC.
        While with $3000\fbm$ data at the LHC, nearly all of the samples can be checked at $5\sigma$ level even if the mass difference is insufficient.
    \item The $a_1$ funnel and the $h_2/Z$ funnel are the two main mechanisms for the singlino-dominated LSP annihilating, which can not be distinguished by searching for NLSPs.

\end{itemize}

\acknowledgments

We thank Yuanfang Yue, Yang Zhang and Liangliang Shang for useful discussions.

\end{multicols}

\vspace{-2.5mm} \centerline{\rule{80mm}{0.1pt}} \vspace{1mm}

\begin{multicols}{2}

\end{multicols}

\clearpage

\end{document}